\documentclass[journal]{IEEEtran} 
\usepackage{setspace}
\usepackage{cite}
\usepackage[pdftex]{graphicx}
\graphicspath{ {./image/} }
\usepackage[cmex10]{amsmath}
\usepackage{dsfont}
\usepackage{array}
\ifCLASSOPTIONcompsoc
 \usepackage[caption=false,font=normalsize,labelfont=sf,textfont=sf]{subfig}
\else
  \usepackage[caption=false,font=footnotesize]{subfig}
\fi

\usepackage{bm}
\usepackage{color}
\usepackage{stfloats}
\usepackage{arydshln} 
\usepackage[]{changes}
\usepackage{url}
\usepackage{textcomp} 
\usepackage{adjustbox} 
\usepackage{textcase}
\usepackage{graphicx}
\usepackage{enumerate}
\usepackage{footnote}
\usepackage{xparse}
\usepackage{amssymb}
\usepackage{amsmath}
\usepackage{amsfonts}
\RequirePackage{xspace}
\RequirePackage{graphics}
\usepackage{xcolor}
\RequirePackage{textcomp}
\usepackage{keyval}
\usepackage{xspace}
\usepackage{mathrsfs}
\usepackage{paralist}
\usepackage{lipsum}
\usepackage{listings}
\usepackage{multirow}
\usepackage{array}
\usepackage{pifont} 
\usepackage{stmaryrd}
\usepackage{textcase}
\usepackage{makecell}
\usepackage{stackengine}%
\usepackage{threeparttable}
\usepackage{balance}
\RequirePackage{tikz}
\usetikzlibrary{arrows,automata,shapes,calc,through,decorations.pathmorphing,decorations.fractals,chains,shapes.multipart}
\usepackage{arydshln} 
\usepackage{etoolbox} 
\usepackage{circuitikz}
\allowdisplaybreaks

\newcommand{\eqlabel}[1]{\label{eq:#1}}
\renewcommand{\eqref}[1]{~(\ref{eq:#1})}

\begin{document}
	
	\title{Multiobjective Backstepping Controller for Parallel Buck Converter}
	
	\author{Ajay~Pratap~Yadav,~\IEEEmembership{Student Member,~IEEE,}
		\thanks{	
			The author is with the Department of Electrical Engineering, University of Texas, Arlington, TX 76019 USA (e-mail: ajay.yadav@mavs.uta.edu)}
	}
	\maketitle
	
	\begin{abstract}
		A backstepping controller is designed for a system of parallel buck converters sharing load. Controller objective is to ensure proper current sharing and output voltage regulation. The designed controller is successfully tested for both constant load and sudden change in loading conditions.
	\end{abstract}

	\begin{IEEEkeywords}
		DC-DC converters, buck converter, backstepping control, Lyapunov Function
	\end{IEEEkeywords}

\IEEEpeerreviewmaketitle

\newcommand{\mysmallarraydecl}{\renewcommand{
\IEEEeqnarraymathstyle}{\scriptscriptstyle}%
\renewcommand{\IEEEeqnarraytextstyle}{\scriptsize}
\settowidth{\normalbaselineskip}{\scriptsize
\hspace{\baselinestretch\baselineskip}}%
\setlength{\baselineskip}{\normalbaselineskip}%
\setlength{\jot}{0.25\normalbaselineskip}%
\setlength{\arraycolsep}{2pt}}

\section{Introduction}
Parallel DC-DC converters are useful in high-performance applications due to their ability to reduce stress on individual components and improve reliability. But these benefits come at a price as controller design becomes more involved. The objective of the controller is to ensure voltage regulation and proper load sharing. If the controller fails to achieve these goals, this can lead to converter overloading and bad output voltage profile which can further damage the system. 

The traditional droop mechanism that is used for load sharing among parallel generators for AC systems can be extended to parallel converters \cite{guerrero2011hierarchical}, \cite{dragicevic2015dc}. Although the droop mechanism is easy to implement, its passive nature has motivated researchers to look for active load sharing methods. The concept of master-slave method is quite popular in the scientific community \cite{grbovic2009master}, \cite{giri2006common}. In this method, one converter acts as a master and regulates the output voltage while the other converters (slaves) follow the master's current. The concept of multi-agent systems has also be applied for parallel DC-DC systems \cite{moayedi2015team, nasirian2014distributed}. In this paper, separate backstepping controllers are designed for the two converters. The first converter aims to follow the given reference voltage while the second converter ensures proportional load sharing. 

\section{System Modeling and Problem Formulation} \label{sec_modeling}
\begin{figure}
	\centering
	\includegraphics[scale=0.38]{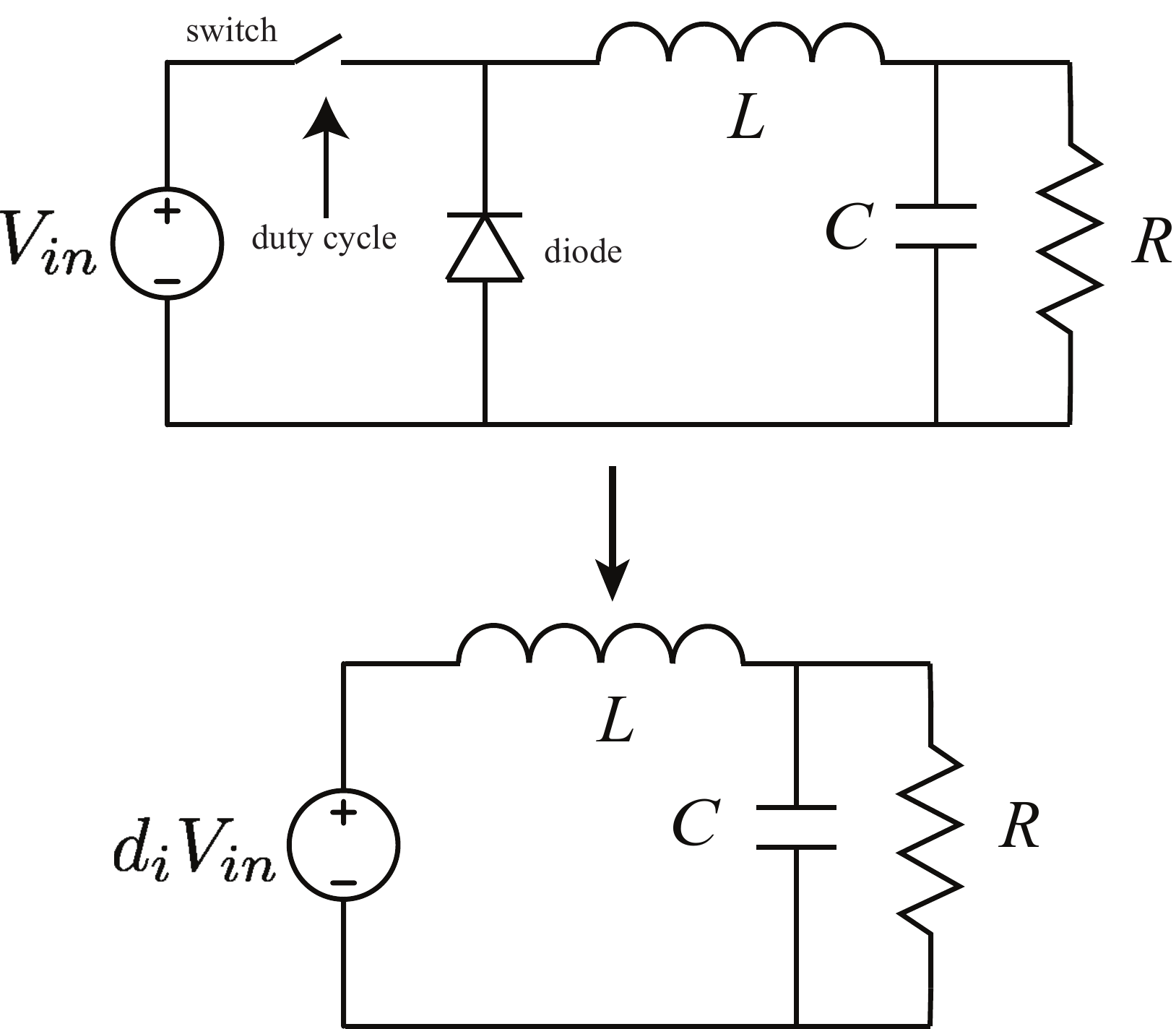}
	\caption{Buck converter and its equivalent model.}
	\label{fig_buck1}
\end{figure}

Figure \ref{fig_buck1} shows an ideal DC-DC buck converter along with its equivalent model which can be obtained using average-value modeling \cite{erickson2001fundament}, \cite{611275}. System parameters for a buck converter are input voltage, $V_{in}$, inductance, $L$, capacitance, $C$ and the load resistance, $R$.   

\begin{figure}
	\centering
	\includegraphics[scale=0.40]{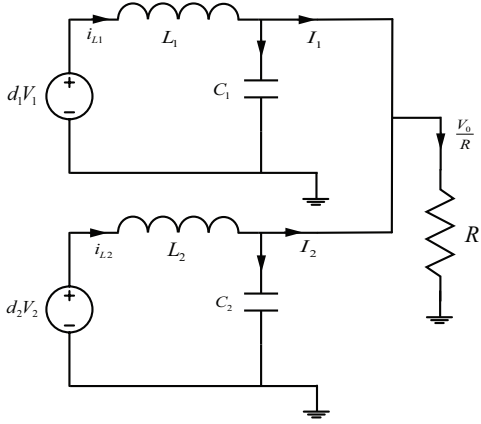}
	\caption{Averaged model of parallel buck converter.}
	\label{fig_buck2}
\end{figure}

For the given buck converter, the control variable is the duty cycle $d_i$, which is the ratio of the on-time of the MOSFET (switch) and the switching time period \cite{erickson2001fundament}. Figure \ref{fig_buck2} shows the system of two parallel buck converters supplying a common load resistance $R$. The state space model for this system can be derived as follows. Appyling the Kirchhoff voltage law on two inductors, we get
\begin{align}
L_1\frac{di_{L_1}}{dt} &= d_1V_1-V_{o} \nonumber \\
L_2\frac{di_{L_2}}{dt} &= d_2V_2-V_{o}
\eqlabel{eq_syst1}
\end{align} 
Applying the Kirchhoff current law gives
\begin{align}
C_1 \frac{dV_{o}}{dt} &= i_{L_1}-I_1 \nonumber \\
C_2 \frac{dV_{o}}{dt} &= i_{L_2}-I_2 \nonumber \\
I_1+I_2 &= V_o/R
\eqlabel{eq_syst2}
\end{align}
Adding the equations \eqref{eq_syst2} will give
\begin{align}
C \frac{dV_{o}}{dt} = i_{L_1}+i_{L_2}-V_o/R
\eqlabel{eq_syst3}
\end{align}     
where, $C = C_1+C_2$. 

The objective  of the control system is to make sure that the output voltage is regulated at a given $V_{ref}$ and the current is shared proportionally between the converters. This means that $V_o \rightarrow V_{ref}$ and $i_{L_1}/I_{1m} = i_{L_2}/I_{2m}$, where $I_{1m}$ and $I_{2m}$ are the rated current for converter 1 and converter 2, respectively. It is assumed that the converters stay in the continuous current mode (CCM). 
\section{Backstepping Control} \label{sec_method1}
\subsection{Voltage Regulation}
The control objectives for this system are achieved by using separate backstepping-based controller for each converter. The first converter ensures that the output voltage of converter 1 traces the given reference voltage, $V_{ref}$.  The second converter adjusts its current such that the load sharing is proportional. 

From equations \eqref{eq_syst1} and \eqref{eq_syst3}, state space equation for Converter 1 is 
\begin{align}
L_1\frac{di_{L_1}}{dt} &= d_1V_1-V_{o} \nonumber \\
C \frac{dV_{o}}{dt} &= i_{L_1}+i_{L_2}-V_o/R
\end{align}  
where $C = C_1+C_2$. 
Let $e = V_{ref}-V_o$, the error between the desired and actual output voltage. Taking the derivative, we get
\begin{align}
\dot{e} &= \frac{1}{C}(\frac{Vo}{R}-i_{L_1}-i_{L_2}) \nonumber \\
\dot{e} &= \frac{1}{C} (\frac{V_{ref}}{R}-\frac{e}{R}-i_{L_1}-i_{L_2})
\eqlabel{eq_back1}
\end{align}
In equation \eqref{eq_back1}, $i_{L_1}$ acts as the virtual control variable. Therefore, adding and subtracting $i_{L_{1D}}$, which is the desired value for $i_{L_1}$ we get
\begin{align}
C\dot{e}+\frac{e}{R}-\frac{V_{ref}}{R}+\frac{i_{L_2}}{R}+i_{L_{1D}} = \tilde{i}_{L_1}
\end{align}
where, $\tilde{i}_{L_1} = i_{L_{1D}}-i_{L_1}$. If $i_{L_{1D}} = \frac{V_{ref}}{R}-\frac{i_{L_2}}{R}+k_1e$, where $k1>0$, the error dynamics becomes 
\begin{align}
C\dot{e}+e\left(\frac{1}{R}+k_1\right)= \tilde{i}_{L_1}
\end{align}
which is a stable system. The variable $k_1$ is added to ensure faster response. 

In order to select the appropriate duty cycle $d_1$, let 
$\dot{i}_{L_1}=W_1$. Consider the Lyapunov Function candidate (LFC)
\begin{align}
V_1 &= \frac{1}{2} \left( e^2+\tilde{i}_{L_1}^2\right) \nonumber \\
\dot{V}_1 &= e\dot{e}+\tilde{i}_{L_1} \dot{\tilde{i}}_{L_1} \nonumber \\
\dot{V}_1 &= \frac{e}{C}\left(\tilde{i}_{L_1}-\frac{e}{R}-k_1e\right)+\tilde{i}_{L_1}\left(-\frac{\dot{i}_{L_2}}{R}+k_1\dot{e}-W_1 \right)
\end{align}
For the Lyapunov function to be negative definite, select 
\begin{align}
W_1 = -\frac{\dot{i}_{L_2}}{R}+\tilde{i}_{L_1}+\frac{e}{C}+k_1 \dot{e}
\eqlabel{eq_W1}
\end{align}
Substituting $W_1$ in Lyapunov function expression gives
\begin{align}
\dot{V}_1 = -\frac{e^2}{C}\left(\frac{1}{R}+k_1 \right)-\tilde{i}_{L_1}^2
\end{align}
Since $\dot{V}_1$ is negative definite, error $e \rightarrow 0$ and $\tilde{i}_{L_1} \rightarrow 0$
Therefore, the duty cycle for the first converter for successful voltage tracking is 
\begin{align}
d_1 = \frac{L_1W_1+Vo}{V_1}
\end{align}
\subsection{Proportional Current Sharing}
The current sharing is ensured by a separate backstepping controller which controls Converter 2. For proportional current sharing between the two converters, say the maximum current that can be supplied by the two converters be $I_{1m}$ and $I_{2m}$. Then the goal of the second controller is to ensure $\left( i_{L_1}/I_{1m} - i_{L_2}/I_{2m} \right) \rightarrow 0$. The state space equation for Converter 2 is as follows
\begin{align}
L_2\frac{di_{L_2}}{dt} &= d_2V_2-V_{o} \nonumber \\
C \frac{dV_{o}}{dt} &= i_{L_1}+i_{L_2}-V_o/R
\eqlabel{eq_system2}
\end{align} 

Let $e_2 = i_{L_1}/I_{1m} - i_{L_2}/I_{2m}$, then the error dynamics for the second controller can be written as
\begin{align}
\dot{e}_2 &= \frac{1}{I_{1m}L_1}(d_1V_1-V_o)-\frac{1}{I_{2m}L_2}(d_2V_2-V_o) \nonumber \\
\dot{e}_2 &= \left(\frac{d_1V_1}{I_{1m}L_1}-\frac{d_2V_2}{I_{2m}L_2}\right)+V_o\left(\frac{1}{I_{2m}L_2}-\frac{1}{I_{1m}L_1} \right)
\eqlabel{eq_back2}
\end{align}
Let $\left(\frac{1}{I_{2m}L_2}-\frac{1}{I_{1m}L_1} \right) = X$, then equation \eqref{eq_back2} becomes 
\begin{align}
\dot{e}_2-\frac{d_1V_1}{I_{1m}L_1}+\frac{d_2V_2}{I_{2m}L_2}-X V_{oD} = -\tilde{V}_o X
\end{align}
where, $V_{oD}$ is the desired output voltage and  $\tilde{V}_o = V_{oD}-V_o$.
If $V_{oD} = \frac{1}{X}\left( -\frac{d_1V_1}{I_{1m}L_1}+\frac{d_2V_2}{I_{2m}L_2}-k_2e_2 \right)$, where $k_2>0$, then the error dynamics becomes a stable system as shown below
\begin{align}
\dot{e}_2+k_2e = -\tilde{V}_o X
\end{align}
To check the stability of the system, consider the Lyapunov Function candidate 
\begin{align}
V_2 &= \frac{1}{2}\left(e_2^2+\tilde{V}_o^2 \right) 
\end{align}
Let $W_2 = \dot{i}_{L_2}$ and differentiate the LFC, which gives
\begin{align}
\dot{V}_2 = &e_2\dot{e}_2+\tilde{V}_o \dot{\tilde{V}}_o \nonumber \\
\dot{V}_2 =  &e_2 \left(-\tilde{V}_o X-k_2e_2 \right) + \nonumber \\
& \tilde{V}_o \left[\frac{1}{X}\left(\frac{V_2\dot{d}_2}{I_{2m}L_2}-k_2\dot{e}_2 \right)-W_2 \right]
\eqlabel{eq_back3}
\end{align}
Note that in equation \eqref{eq_back3}, since the controller is designed considering converter 2, the duty cycle for converter 1, $d_1$, is treated constant. 
For equation \eqref{eq_back3} to be negative definite, select 
\begin{align}
W_2 = \frac{1}{X}\left(\frac{V_2\dot{d}_2}{I_{2m}L_2} - k_2\dot{e}_2 \right)-e_2 X+\tilde{V}_o
\eqlabel{eq_back4}
\end{align}
Since $W_2 = \dot{i}_{L_2}$, using equations \eqref{eq_back4} and \eqref{eq_system2} we get 
\begin{align}
\dot{d}_2 = \frac{I_{2m}L_2}{V_2} \left[ \left(\dot{V}_o+eX-\tilde{V}_o \right)X+k_2\dot{e}_2 \right]
\end{align}
\section{Simulation Results} \label{sec_results}
\subsection{Constant Load}
The designed controller is first tested on a constant resistor load. The system parameters for each converter is given in Table \ref{tab_table1}. The two converters are similar but the maximum current capacity of converter 1 is 5 A while that of converter 2 is 2 A. The parallel converters are supplying current to a 10 $\Omega$ resistor. The goal is to ensure that the per unit current supplied by each converter is equal and the output voltage is regulated at a given $V_{ref}$. Herein, $V_{ref} = 8$, $k_1=k_2=1$. 

\begin{figure}
	\centering
	\includegraphics[width=\linewidth]{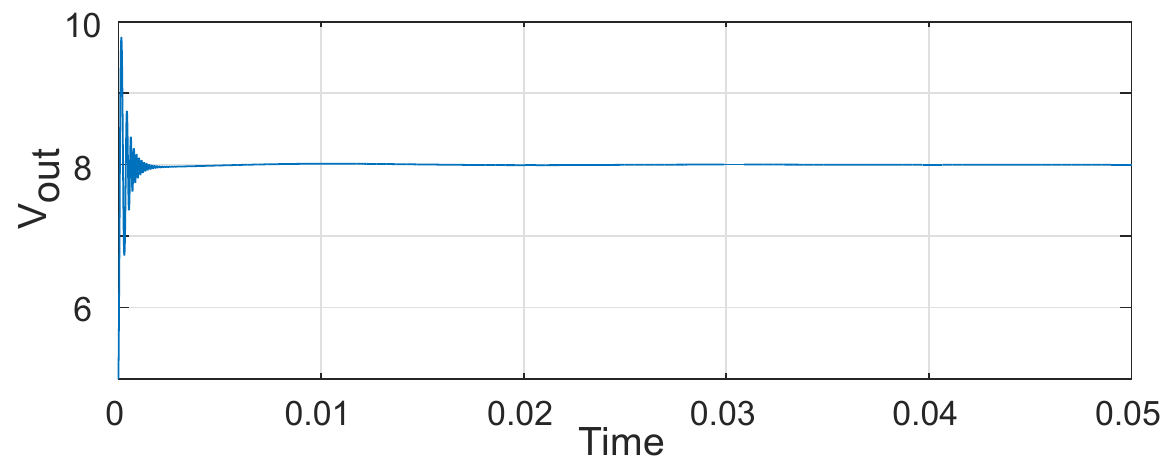}
	\caption{Output voltage for parallel buck converter system. $V_{out}$ is regulated to the reference voltage. }
	\label{fig_V1}
\end{figure}

\begin{figure}
	\centering
	\includegraphics[width=\linewidth]{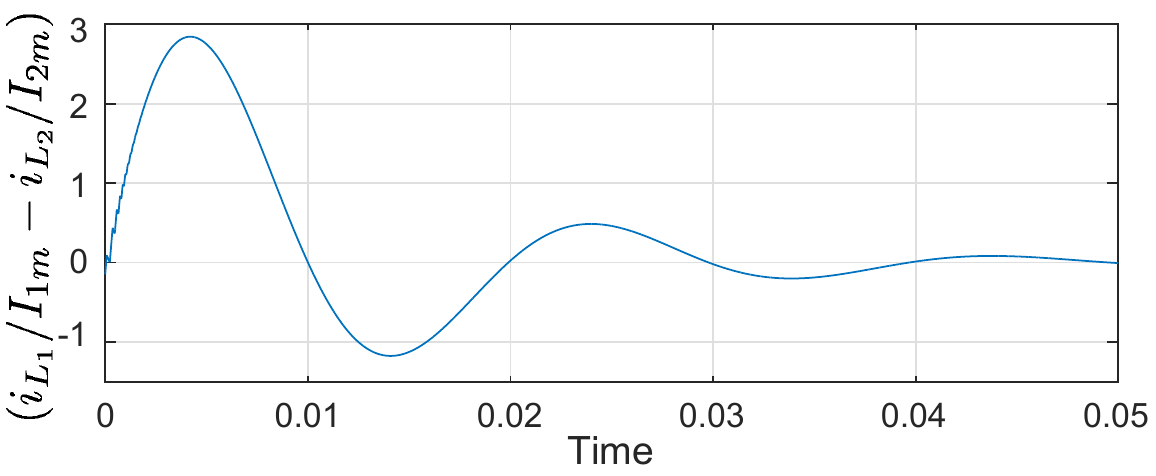}
	\caption{Difference between the per unit currents supplied by each converter. The controller ensures proportional load sharing as seen by the difference in per unit currents. }
	\label{fig_I_diff}
\end{figure}

Figure \ref{fig_V1} shows the variation of output voltage. It can be seen that the controller matches the output voltage to the desired voltage. Figure \ref{fig_I_diff} shows the plot for the difference between the per unit current supplied by both the sources. It can be observed that the controller ensures that the per unit current supplied by each converter is same. 
\begin{table}
	\caption{System Parameters}
	\centering
\scalebox{0.8}{
	\begin{tabular}{|c | c | c |c |}
		\hline
		& $L_i(mH)$ & $C_i(\mu F)$ & $I_{1m}(A)$ \\
		\hline 
		Converter 1 & 1 & 10 & 5 \\
		\hline 
		Converter 2 & 1 & 10 & 2\\ 
		\hline 
	\end{tabular} }
	\label{tab_table1}
\end{table}

\subsection{Change in Resistance}
\begin{figure}
	\centering
	\includegraphics[width=\linewidth]{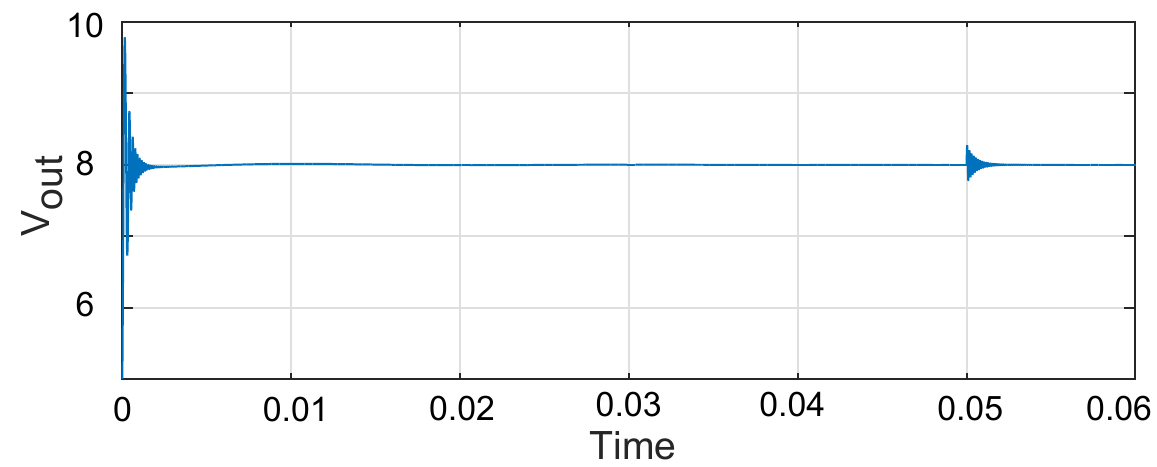}
	\caption{Output voltage $V_{out}$ under sudden change in load. }
	\label{fig_V12}
\end{figure}

\begin{figure}
	\centering
	\includegraphics[width=\linewidth]{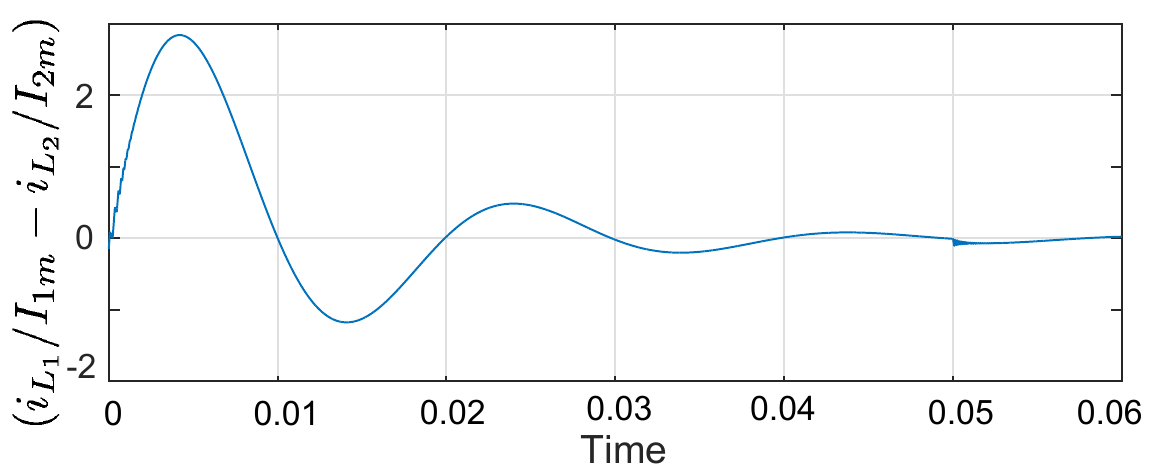}
	\caption{Difference between the per unit currents supplied by each converter.}
	\label{fig_I_diff2}
\end{figure}

In this section, performance of the controller is evaluated when there is a sudden change in load. For this simulation, the resistance of the load is suddenly changed from 10 $\Omega$ to 15 $\Omega$ at time $t=0.05$ seconds. Figures \ref{fig_V12} and \ref{fig_I_diff2} show the variation of output voltages and the difference between the per unit currents. It can be seen that controller quickly regulates the output voltage and balances the current sharing. 
\section{Conclusion}
In this paper, backstepping-based controller is designed for the system of parallel converter sharing a common load. The controller is designed such that the output voltage is regulated and the current is shared proportionally between the converters.

\bibliographystyle{IEEEtran}
\bibliography{References}

\begin{thebibliography}{1}
\providecommand{\url}[1]{#1}
\csname url@samestyle\endcsname
\providecommand{\newblock}{\relax}
\providecommand{\bibinfo}[2]{#2}
\providecommand{\BIBentrySTDinterwordspacing}{\spaceskip=0pt\relax}
\providecommand{\BIBentryALTinterwordstretchfactor}{4}
\providecommand{\BIBentryALTinterwordspacing}{\spaceskip=\fontdimen2\font plus
\BIBentryALTinterwordstretchfactor\fontdimen3\font minus
  \fontdimen4\font\relax}
\providecommand{\BIBforeignlanguage}[2]{{%
\expandafter\ifx\csname l@#1\endcsname\relax
\typeout{** WARNING: IEEEtran.bst: No hyphenation pattern has been}%
\typeout{** loaded for the language `#1'. Using the pattern for}%
\typeout{** the default language instead.}%
\else
\language=\csname l@#1\endcsname
\fi
#2}}
\providecommand{\BIBdecl}{\relax}
\BIBdecl

\bibitem{guerrero2011hierarchical}
J.~M. {Guerrero}, J.~C. {Vasquez}, J.~{Matas}, L.~G. {de Vicuna}, and
  M.~{Castilla}, ``Hierarchical control of droop-controlled ac and dc
  microgrids—a general approach toward standardization,'' \emph{IEEE Trans.
  Ind. Electron.}, vol.~58, no.~1, pp. 158--172, Jan. 2011.

\bibitem{dragicevic2015dc}
T.~{Dragicevic}, X.~{Lu}, J.~C. {Vasquez}, and J.~M. {Guerrero}, ``Dc
  microgrids—part i: A review of control strategies and stabilization
  techniques,'' \emph{IEEE Trans. Power Electron.}, vol.~31, no.~7, pp.
  4876--4891, Jul. 2016.

\bibitem{grbovic2009master}
P.~J. {Grbovic}, ``Master/slave control of input-series- and
  output-parallel-connected converters: Concept for low-cost high-voltage
  auxiliary power supplies,'' \emph{IEEE Trans. Power Electron.}, vol.~24,
  no.~2, pp. 316--328, Feb. 2009.

\bibitem{giri2006common}
R.~{Giri}, V.~{Choudhary}, R.~{Ayyanar}, and N.~{Mohan}, ``Common-duty-ratio
  control of input-series connected modular dc-dc converters with active input
  voltage and load-current sharing,'' \emph{IEEE Trans. Ind. Appl.}, vol.~42,
  no.~4, pp. 1101--1111, Jul. 2006.

\bibitem{moayedi2015team}
S.~{Moayedi}, V.~{Nasirian}, F.~L. {Lewis}, and A.~{Davoudi}, ``Team-oriented
  load sharing in parallel dc–dc converters,'' \emph{IEEE Trans. Ind. Appl.},
  vol.~51, no.~1, pp. 479--490, Jan. 2015.

\bibitem{nasirian2014distributed}
V.~{Nasirian}, A.~{Davoudi}, and F.~L. {Lewis}, ``Distributed adaptive droop
  control for dc microgrids,'' in \emph{Proc. 29th IEEE Appl. Power Electron.
  Conf. Expo.}, 2014, pp. 1147--1152.

\bibitem{erickson2001fundament}
R.~W. Erickson and D.~Maksimovic, \emph{Fundament of Power Electronics},
  2nd~ed.\hskip 1em plus 0.5em minus 0.4em\relax New York, USA: Kluver Academic
  Publishers, 2001.

\bibitem{611275}
J.~Mahdavi, A.~Emaadi, M.~D. Bellar, and M.~Ehsani, ``Analysis of power
  electronic converters using the generalized state-space averaging approach,''
  \emph{IEEE Trans. Circuits Syst. I: Fundam. Theory Appl.}, vol.~44, no.~8,
  pp. 767--770, Aug 1997.

\end{thebibliography}

\end{document}